**Extraordinary nonlinear optics in ordinary semiconductors**

N. Akozbek[1], V. Roppo[1,2], M.A. Vincenti[1,3], J.V. Foreman[1], M.J. Bloemer[1], J.W. Haus[4], M. Scalora[1]

[1]*Charles M. Bowden Research Center, AMSRD-AMR-WS-ST, RDECOM, Redstone Arsenal, Alabama 35898-5000, USA*

[2]*Departament de Fisica i Enginyeria Nuclear, Universitat Politecnica de Catalunya C/Colom 11, 08222 Terrassa Spain*

[3]*Dipartimento di Elettrotecnica ed Elettronica, Politecnico di Bari, Via Orabona 4, 70125 Bari, Italy*

[4]*Electro-Optics Program, University of Dayton, Dayton, OH 45469-0245, USA*

**Abstract**

We numerically demonstrate inhibition of absorption, optical transparency, and anomalous momentum states of phase locked harmonic pulses in semiconductors, at UV and extreme UV frequencies, in spectral regions where the dielectric constant of typical semiconductors is negative. We show that a generated harmonic signal can propagate through a bulk metallic medium without being absorbed as a result of a phase locking mechanism between the pump and its harmonics. These findings may open new regimes in nonlinear optics and are particularly relevant to the emerging fields of nonlinear negative index meta-materials and nano-plasmonics, especially in the ultrafast pulse regime.

High harmonic generation is a fundamental process in nonlinear optics that has been used to generate new coherent light sources towards shorter wavelengths (for a recent review see [1]). Compact coherent sources particularly at extreme ultraviolet (EUV) spectral region have important applications ranging from nanotechnology, high resolution microscopy and spectroscopy [1]. Most nonlinear optical materials absorb strongly in the UV region (~<200nm) and harmonic generation has been usually limited to visible wavelengths in order to avoid absorption losses. Absorption at the harmonic wavelength is usually considered detrimental because it is thought to limit the conversion efficiency over the coherence length. In particular, in cases where the sample is much longer than the characteristic absorption length and coherence length it is expected that any generated harmonics will be completely re-absorbed inside the medium. However, the generation and transmission of second and third harmonics in a bulk GaAs was recently demonstrated both theoretically and experimentally, in a case where the harmonic frequencies were tuned well above the absorption band edge [2]. A 100 fs pump pulse was tuned at 1300nm and the generated second and third harmonics at 650nm and 435 nm respectively propagated through a 450 micron GaAs sample, where the absorption edge is around 890nm. This extraordinary phenomenon is attributed to phase locking between the fundamental and its harmonics and occurs under phase mismatched conditions irrespective of material parameters [3]. As a result of the phase locking mechanism, the pump impresses its dispersive properties to the generated harmonics which as a consequence will experience no absorption or other de-phasing effects such as group velocity walk-off. The harmonics then co-propagate locked with the pump and any energy exchange between the pump and its harmonics is clamped. In transparent materials, the general solution for second harmonic generation from a boundary layer is composed of a reflected signal and two forward propagating components, one displaying a k-vector that is the solution of the homogenous wave equation $k_{SH} = k_{2\omega}$ and the other that is a solution of the inhomogeneous wave equation, equal to twice the pump wave vector $k_{SH} = 2k_{\omega}$ [4]. For example, Maker fringes are the result of the interference between the two forward propagating components that at phase matching merge into a single solution [5]. Recently, complete phase and group velocity locking were demonstrated in a Lithium Niobate wafer, for different combinations of incident polarizations [6].

In this letter, we demonstrate second harmonic generation in a new regime where the dielectric permittivity is negative. This regime is of great interest in the development of negative index materials and nano-plasmonic optical devices. Bulk metals have their plasma frequencies in the UV-VIS region due to their relatively large free carrier densities, and exhibit a negative permittivity from the visible to microwave frequencies. Unlike bulk metals, semiconductors such as GaAs, GaP and Ge exhibit a region of negative dielectric permittivity in the deep UV, while being transparent in the infrared region. Negative refraction that is in general associated with negative index materials ($\varepsilon<0$, and $\mu<0$) occurs also for a TM-polarized field incident on a negative permittivity medium [7]. This has led to the development of a tunable superlens in the visible range using resonant metal/dielectric multilayer structures [8]. The same arrangement can be exploited in resonant multilayer stacks such as GaAs/KCL and GaAs/MgO operating in the deep UV spectral region [9]. These structures may provide an important step in the experimental realization of a superlens in the optical regime, since the fabrication of bulk, effective negative index materials has proven to be very challenging due to relatively large absorption losses. Losses are still a major limiting factor for any practical applications. Semiconductors have been extensively used in photonics as well as electronics industry and thus are easier to integrate with current electro-optical technologies. Unlike bulk metals, semiconductors can be doped to control free carrier densities and the free carrier plasma frequency. In fact, doped semiconductors that exhibit a negative permittivity in the far IR region were shown to support negative refraction [10], plasmonic response [11], and superlensing in SiC [12]. In addition, semiconductors exhibit large nonlinear coefficients, an important factor for nonlinear frequency conversion and other types of parametric processes.

The phase locking mechanism reported in [3] is responsible for inhibiting the absorption of generated 650nm (second harmonic) and 435nm (third harmonic) light from a GaAs substrate pumped at 1300nm. Note that both harmonic wavelengths are well below the absorption edge (~890nm), but still in the range of normal material dispersion (Re[ε]>0) [2, 13]. However, the real part of ε of GaAs, for instance, is negative between approximately 120nm and 262nm (Inset Fig.1, shaded area) [13]. Some questions that naturally arise are as follows: Does the phase locking mechanism still manifest itself in the metallic region, where all propagating modes should naturally be suppressed? If so, can absorption still be inhibited when the pump is tuned in such a way that harmonic generation occurs in the negative ε region? Could anomalous momentum states of the type already reported for harmonic generation in pure metamaterials (ε<0 and μ<0) [14] exist in this regime (ε<0, μ=1), using TM-polarized waves in ordinary semiconductors? Our results suggest that: (1) the inhibition of absorption is not limited by the spectral position of the harmonics or proximity of a resonance [2], nor is it constrained by the relative or absolute signs of ε and μ: the mechanism is triggered even in metamaterials where ε and μ are simultaneously negative. A sufficient condition for the onset of the inhibition of absorption appears to be that the pump be tuned within a range of transparency; (2) anomalous momentum states manifest themselves in the metallic wavelength range of semiconductors, yielding fields that effectively propagate inside bulk metallic substrates.

Instead of presenting theoretical results specific to a particular semiconductor, in what follows we undertake a general discussion of the dynamical aspects of nonlinear pulse propagation phenomena across an absorption resonance, where the dielectric constant takes on positive and negative values, for a pump pulse tuned to a range of transparency. For simplicity, assume that the material response can be modeled by a collection of singly-resonant Lorentz oscillators. To model second and/or higher harmonic generation we assume that a TM-polarized field may be decomposed as a superposition of harmonics as follows [2]:

$$\mathbf{H} = \hat{\mathbf{x}} \sum_{\ell=1}^{\infty} \left( H_x^{\ell\omega}(z,y,t) + c.c \right) = \hat{\mathbf{x}} \sum_{\ell=1}^{\infty} \left( \mathcal{H}_x^{\ell\omega}(z,y,t) e^{i\ell(kz-\omega t)} + c.c \right)$$

$$\mathbf{E} = \hat{\mathbf{y}} \sum_{\ell=1}^{\infty} \left( E_y^{\ell\omega}(z,y,t) + c.c \right) + \hat{\mathbf{z}} \sum_{\ell=1}^{\infty} \left( E_z^{\ell\omega}(z,y,t) + c.c \right) =$$

$$\hat{\mathbf{y}} \sum_{\ell=1}^{\infty} \left( \mathcal{E}_y^{\ell\omega}(z,y,t) + c.c \right) + \hat{\mathbf{z}} \sum_{\ell=1}^{\infty} \left( \mathcal{E}_z^{\ell\omega}(z,y,t) + c.c \right)$$
, (1)

where $\mathcal{E}_y^{\ell\omega}(z,y,t)$, $\mathcal{H}_y^{\ell\omega}(z,y,t)$ are generic, spatially- and temporally-dependent, complex envelope functions; $k$ and $\omega$ are carrier wave vector and frequency, respectively, and $\ell$ is an integer. Eqs.(1) are a convenient representation of the fields, and no *a priori* assumptions are made about the envelopes. For simplicity, we assume a TM-polarized incident field and similarly polarized harmonics. The linear response of the medium is described by a Lorentz oscillator model: $\varepsilon(\omega) = 1 - \frac{\omega_p^2}{\omega^2 + i\gamma\omega - \omega_r^2}$, and $\mu(\omega) = 1$, where $\gamma$, $\omega_p$, and $\omega_r$ are the damping coefficient, the plasma and resonance frequencies, respectively. Second order nonlinearities are introduced as a nonlinear polarization of the type: $P_{NL} = \chi^{(2)} E^2$. Retaining terms up to the second harmonic field, and allowing for a dispersive nonlinear coefficient, the corresponding nonlinear polarization terms are $\mathcal{P}_\omega(z,t) = 2\chi_\omega^{(2)} \mathcal{E}_\omega^* \mathcal{E}_{2\omega}$ and $\mathcal{P}_{2\omega}(z,t) = \chi_{2\omega}^{(2)} \mathcal{E}_\omega^2$. Assuming that polarization and currents may be decomposed as in Eqs.(1), we obtain the following Maxwell-Lorentz system of equations for the $\ell^{th}$ field components, in a two-dimensional space ($\tilde{y}, \xi$) plus time ($\tau$) coordinate system:

$$\frac{\partial \mathcal{H}_x^{\ell\omega}}{\partial \tau} = i\beta_\ell \left( \mathcal{H}_x^{\ell\omega} + \mathcal{E}_z^{\ell\omega} \sin\theta_i + \mathcal{E}_y^{\ell\omega} \cos\theta_i \right) - \frac{\partial \mathcal{E}_z^{\ell\omega}}{\partial \tilde{y}} + \frac{\partial \mathcal{E}_y^{\ell\omega}}{\partial \xi}$$

$$\frac{\partial \mathcal{E}_y^{\ell\omega}}{\partial \tau} = i\beta_\ell \left( \mathcal{E}_y^{\ell\omega} + \mathcal{H}_x^{\ell\omega} \cos\theta_i \right) - 4\pi(\mathcal{J}_y^{\ell\omega} - i\beta_\ell \mathcal{P}_y^{\ell\omega}) + i4\pi\beta_\ell \mathcal{P}_{y,NL}^{\ell\omega} - 4\pi \frac{\partial \mathcal{P}_{y,NL}^{\ell\omega}}{\partial \tau} + \frac{\partial \mathcal{H}_x^{\ell\omega}}{\partial \xi}$$

$$\frac{\partial \mathcal{E}_z^{\ell\omega}}{\partial \tau} = i\beta_\ell \left( \mathcal{E}_z^{\ell\omega} + \mathcal{H}_x^{\ell\omega} \sin\theta_i \right) - 4\pi(\mathcal{J}_z^{\ell\omega} - i\beta_\ell \mathcal{P}_z^{\ell\omega}) + i4\pi\beta_\ell \mathcal{P}_{z,NL}^{\ell\omega} - 4\pi \frac{\partial \mathcal{P}_{z,NL}^{\ell\omega}}{\partial \tau} - \frac{\partial \mathcal{H}_x^{\ell\omega}}{\partial \tilde{y}} \quad . \quad (2)$$

$$\frac{\partial \mathcal{J}_y^{\ell\omega}}{\partial \tau} = \left(2i\beta_\ell - \gamma\right) \mathcal{J}_y^{\ell\omega} + \left(\beta_\ell^2 + i\gamma\beta_\ell - \beta_r^2\right) \mathcal{P}_y^{\ell\omega} + \frac{\pi\omega_p^2}{\omega_0^2} \mathcal{E}_y^{\ell\omega}$$

$$\frac{\partial \mathcal{J}_z^{\ell\omega}}{\partial \tau} = \left(2i\beta_\ell - \gamma\right) \mathcal{J}_z^{\ell\omega} + \left(\beta_\ell^2 + i\gamma\beta_\ell - \beta_r^2\right) \mathcal{P}_z^{\ell\omega} + \frac{\pi\omega_p^2}{\omega_0^2} \mathcal{E}_z^{\ell\omega}$$

$$\mathcal{J}_y^{\ell\omega} = \frac{\partial \mathcal{P}_y^{\ell\omega}}{\partial \tau}; \qquad \mathcal{J}_z^{\ell\omega} = \frac{\partial \mathcal{P}_z^{\ell\omega}}{\partial \tau}$$

In Eqs.(2), the functions $\mathcal{J}, \mathcal{P}, \mathcal{P}_{NL}$ refer to linear electric currents, polarization, and nonlinear polarization, respectively. We have scaled the coordinates so that $\xi = z/\lambda_0$, $\tilde{y} = y/\lambda_0$, $\tau = ct/\lambda_0$, $\omega_0 = \frac{2\pi c}{\lambda_0}$, where $\lambda_0 = 1\mu m$ is a convenient reference wavelength; $\gamma$, $\beta_{\ell\omega} = 2\pi\ell\omega/\omega_0$, $\beta_r = 2\pi\omega_r/\omega_0$, $\omega_p$, are the scaled damping coefficient, wave-vector, resonance and electric plasma frequencies for the $\ell^{th}$ harmonic, respectively. $\theta_i$ is the angle of incidence of the pump field. The equations are solved using a split-step, fast Fourier transform-based pulse propagation algorithm [2, 14, 15] that advances the fields in time.

In Fig.1 we plot the complex dielectric function for the Lorentz oscillator described by the following parameters for the second harmonic signal: $\tilde{\gamma} = 0.1$, $\tilde{\omega}_p = 4$, and $\tilde{\omega}_r = 4$, and similar parameters for the pump. The pump is tuned away from the resonance, and losses can be neglected. These choices result in an absorption resonance centered at λ=250nm, and yield a negative dielectric constant between 175nm and 250nm. For illustration purposes, the inset of Fig.1 shows the complex dielectric constant of GaAs [13]. As a representative example we tune a 50-femtosecond pump pulse to 410nm, so that $\varepsilon(410nm) = 2.592 + i\,3.86\times10^{-6}$, and launch it at -25° angle of incidence, as shown in Fig.2.

The SH signal is generated such that $\varepsilon(205nm) = -1.044 + i\,0.128$. The results of our simulation, shown in Fig.2, predict that a phase-locked SH pulse is generated in a wavelength region where the dielectric constant is negative, and is transmitted by a 30μm-thick bulk sample. We note that sample thickness is not important here [2], and that only the phase locked component is generated. The homogeneous component experiences a metallic region and is quickly absorbed-reflected at the surface. In contrast, the phase locked component is generated at the surface and survives because it experiences the dispersion of the pump [2,6], which is tuned within a region of optical transparency. Fig.2b shows that the SH pulse overlaps the pump pulse – Fig.2a – at all times, in this case with a refraction angle of approximately -15.21°.

One question still remains unanswered relative to the manifestation of anomalous momentum states of the type already discussed in reference [14], where phase locking was investigated in media having simultaneously negative ε and μ (propagating wave regime). In reference [14], it was found that under phase mismatched conditions and through phase locking, the pump pulse generates and captures a SH pulse that in response manifests a negative momentum as the pulse travels in the positive direction. The SH pulse is trapped by the pump, so that the pump-SH pulse system may best be described as a bound state, with a total positive electromagnetic momentum. Does the same occur in the case of semiconductors, i.e. realistic materials? How does the negative dielectric constant manifest itself under these conditions, if at all?

In order to answer these questions we once again turn our attention to Fig.2, and calculate the refraction angles of both signals using the components of the electromagnetic momentum on the $(\tilde{y}, \xi)$ plane. The electromagnetic momentum of a wave packet inside a medium of length $L$ can be calculated as: $P_\xi(\tau) = \dfrac{1}{c^2} \int_{\xi=0}^{\xi=L} \int_{\tilde{y}=-\infty}^{\tilde{y}=\infty} S_\xi(\tilde{y},\xi,\tau,\omega)\,d\tilde{y}\,d\xi$, and

$$P_{\tilde{y}}(\tau) = \frac{1}{c^2} \int_{\xi=0}^{\xi=L} \int_{\tilde{y}=-\infty}^{\tilde{y}=\infty} S_{\tilde{y}}(\tilde{y},\xi,\tau,\omega) d\tilde{y} d\xi$$, where $S_{\xi,\tilde{y}}(\tilde{y},\xi,\tau,\omega)$ are the components of the total Poynting vector $\mathbf{S} = \frac{c}{4\pi}\mathbf{E} \times \mathbf{H}$ at each frequency, from which the total electromagnetic momentum may be calculated. Using these components one may extract the momentum refraction angle of each individual wave packet as a function of time inside the medium as: $\theta_r(\tau) = \tan^{-1}\left[P_{\tilde{y}}(\tau)/P_\xi(\tau)\right]$. The dynamics may then be followed to steady state by observing the final values of the momentum components and the corresponding angle, once the wave packet is completely inside the medium. This approach typically yields the same results predicted by Snell's law even for few-cycle wave packets [14]. In the case depicted in Fig.2b, the calculated momentum refraction angle for the SH field is approximately 18.5º, i.e. in the upper right quadrant, and is exemplified by the orange arrow affixed to the SH wave packet moving inside the medium. However, the SH pulse is trapped by the pump and in fact follows the pump's trajectory at -15.92º.

The results depicted in Fig.2 seem quite dramatic, and stand witness to the remarkable characteristics of phase locking and the consequences that it can impart to dynamical systems. The figure shows the theoretical prediction of a harmonic pulse that propagates in a *bulk metallic medium*, without being absorbed. In Fig.3 we show the refraction angles as a function of wavelength, calculated for the pump and the SH pulses using the components of the electromagnetic momentum. The momentum refraction angle for the *pump* (right axis) varies between -16.4º at 560nm (SH at 280nm), and -12.9º at 320nm (SH at 160nm). These values are in excellent agreement with the predictions of Snell's law. On the other hand, the momentum refraction angle of the SH signal (left axis) is obviously not in accord with Snell's law, because the generated SH pulse acquires the effective dispersion experienced by the pump. The result is a SH wave packet that propagates inside a forbidden wavelength range, where the material displays metallic properties, without being absorbed, as depicted on Fig.2.

In Fig.3 we also depict the directions of **S**$_{2\omega}$ and **k**$_{2\omega}$ (wave vector) for the SH signal as a function of wavelength (the **k**-vector is calculated as an expectation value as follows: $<\mathbf{k}(\omega,\tau)> = \int_{k_\xi=-\infty}^{k_\xi=\infty}\int_{k_{\tilde{y}}=-\infty}^{k_{\tilde{y}}=\infty} \left(\hat{\mathbf{y}}k_{\tilde{y}} + \hat{\mathbf{z}}k_\xi\right)|H(\omega,k_{\tilde{y}},k_\xi,\tau)|^2\, dk_{\tilde{y}}dk_\xi$ ). If an attempt is made to analyze the SH pulse separately from the pump, then conditions are reminiscent of an anomalous situation, where it seems that the direction of momentum flow (Fig.2) is discordant with the direction of propagation [11]. While phase locking ensures that the wave vector will point in the same direction as the pump's wave vector (at the pump' Snell angle), the Poynting vector can change direction in dramatic fashion. For example, at long wavelengths ($\lambda$>250nm, $\varepsilon$>0) **S**$_{2\omega}$ points in the lower, right quadrant, just as the pump does, and normal propagation/refraction ensues. However, once the dielectric constant becomes negative ($\lambda$<250nm), the direction of **S**$_{2\omega}$ shifts to the upper right quadrant. This occurs because the SH (TM-polarized) field must satisfy a boundary condition that changes the sign of the longitudinal components of the electric field and redirects the Poynting vector upward, just as it occurs in metal layers [7]. The refraction angle of the SH signal steadily increases until $\varepsilon > 0$ once again ($\lambda$~175nm), where one expects the direction of the Poynting vector to return to the lower right quadrant. However, we note that the precipitous drop of the refraction angle culminates at the bottom of the curve, near 175nm, with a refraction angle of ~-96°, as indicated in the figure.

The existence of anomalous SH momentum states can be easily reconciled with Poynting's theorem, which in fact gives information about total momentum and energy for the bound pump-SH system. Under the conditions we have explored, phase mismatching ensures that the pump remains undepleted for all reasonable nonlinear coefficients and incident peak powers, so that the momentum of the SH signal is always several orders of magnitude smaller compared to the pump momentum. In this view, even though the SH moves in one direction and its momentum points in another, the total momentum and total energy for the bound state always nearly coincide with the pump's momentum and energy. In other words, under realistic conditions the generated SH pulse does not gain enough momentum/energy to significantly affect the direction of motion of the system, at least in the regime under consideration. The situation is summarized in Fig.4. We note that the SH pulse could not exist inside the medium without the pump, and it is remarkable to see that the signal still recognizes the medium as having a negative dielectric permittivity. Even though the phase locking mechanism is valid for TE and TM modes alike, TE waves cannot refract negatively in metallic media: the SH momentum closely tracks the pump momentum and there are no apparent anomalies. Finally, if a SH pulse is seeded from any direction, a small portion of it is always captured by the pump, and apparent anomalies can ensue for both TE- and/or TM-polarized beams, since the phase locking mechanism is triggered in all cases. While we have restricted our analysis to only second harmonic generation these findings are also valid for third and higher harmonics.

In summary, one of the unique properties of semiconductors is that they exhibit a spectral range where the dielectric permittivity is negative, while providing a transparency region at near IR and longer pump wavelengths. Therefore, these results can easily be realized in typical bulk semiconductors. For example, GaP is a potential candidate transparent above 500nm. While the second harmonic pulse might be tuned around the 330nm resonance of GaP, the third harmonic pulse would be tuned at the negative permittivity region. Another feature of semiconductors is that they can be doped to tune the dielectric properties [17]. Also, it has been shown recently that nano-rings could be used to realize metamaterials in the optical regime [18]. Another approach that has been proposed involves the use of a four-level atomic medium having electric and magnetic transitions in hydrogen and neon atoms [19]. Since semiconductors already exhibit an intrinsic negative permittivity, the introduction of suitable dopant atoms could make it possible to use magnetic transitions to create an effective negative magnetic permeability. The nonlinear coefficient of GaAs below the band edge takes on values of order 500pm/V [20]. If successful, this approach would represent a paradigm shift in the quest for a true, practical, bulk, even tunable "negative index" metamaterial. The present effort thus represents the beginning of a larger, more comprehensive study that will serve to bridge the gap between a number of physical effects, such as high harmonic generation, sub-diffraction limited imaging, negative refraction, and surface plasmons using nonlinear semiconductor materials.

**Figurer Captions**

**Fig.1** Generic Lorentz resonance. The parameters are $\omega_0=4$, $\omega_p=4$, $\gamma=10^{(-5)}$ for the pump, and $\omega_0=4$, $\omega_p=4$, $\gamma=0.1$ for the SH field. Inset: Dielectric constant of GaAs, taken from reference [13]. The regions of negative dielectric constant are denoted by the shaded areas.

**Fig.2** (a) A TM-polarized pump pulse ~50fs in duration, tuned to 410nm, (SH at 205, where Re($\varepsilon$)<0) crosses a ~30micron thick medium. According to the Lorentz model of Fig.1, the index of refraction for the pump field is n=1.56. (b) Generated SH signal. Part of the signal is reflected specularly, while part is transmitted. Since $\varepsilon<0$ and $\mu=1$ at the SH wavelength, conventional wisdom suggests no propagating waves are allowed inside the material. However, the SH pulse phase locks to the pump, experiences the same effective dispersion as the pump, and it is able to tunnel through the sample regardless of sample thickness.

**Fig.3** Pump momentum refraction angle (dashed black curve, right axis), and SH momentum refraction angle (dashed red curve, left axis). The pump's momentum refraction angle coincides with predictions made using Snell's law. The SH momentum refraction angle takes on positive and negative values, depending on the sign of the dielectric constant, and oscillates with large amplitude near the ε=0 crossing point (~175nm). In reality the SH pulse becomes trapped and moves in the same direction as the pump.

**Fig.4** Phase locked fundamental (red shell) and SH (blue shell) pulses. The pump momentum and energy nearly always overwhelm corresponding SH values, so that the total momentum of the system (in green) is always approximately equal to the pump momentum. Here we depict a case where for ε<0 the SH momentum points in the upward right quadrant. The total momentum of the system can be affected by larger SH conversion efficiencies, and beam deflections can reach a fraction of degree, which can become noticeable for large propagation distances.

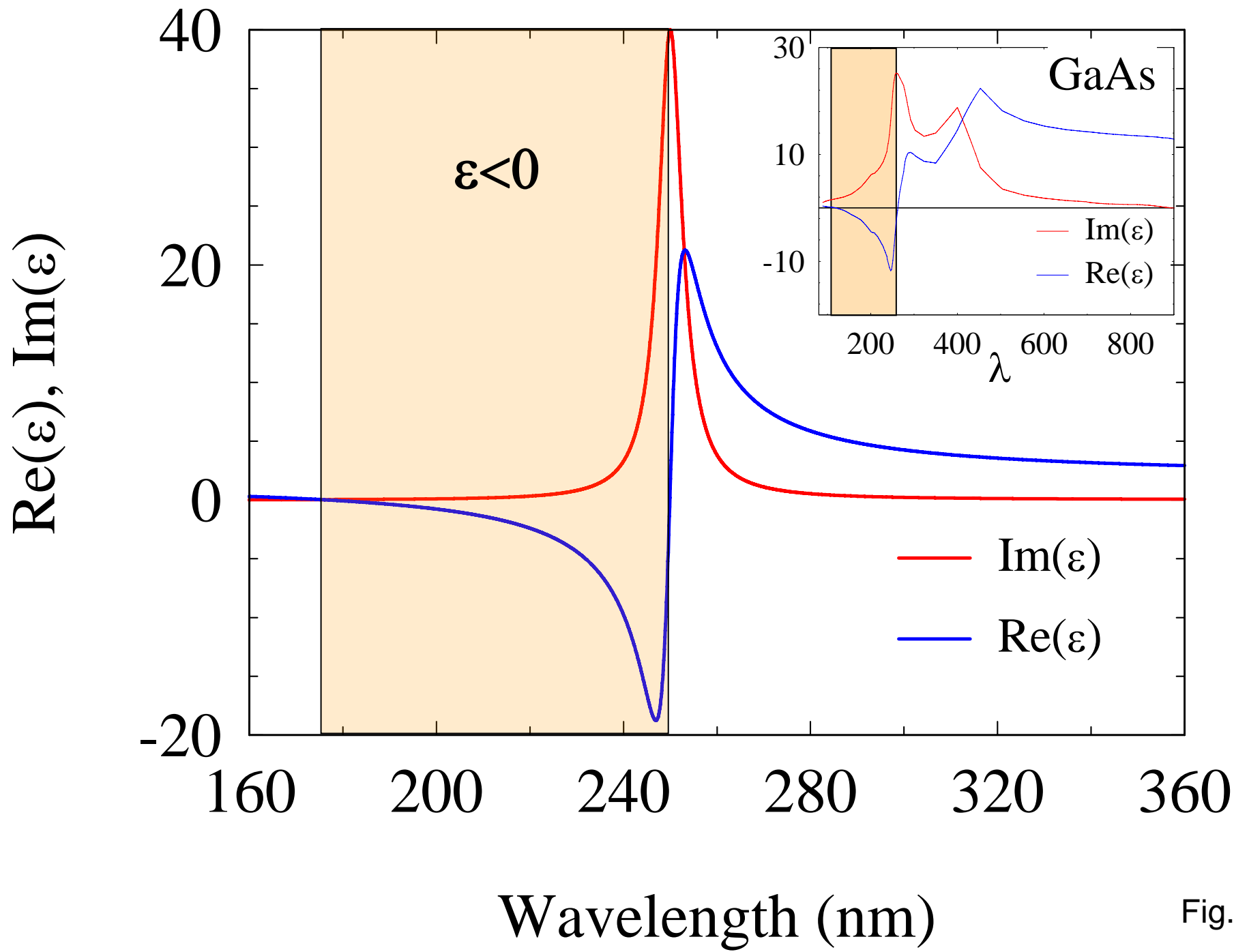

Fig.1

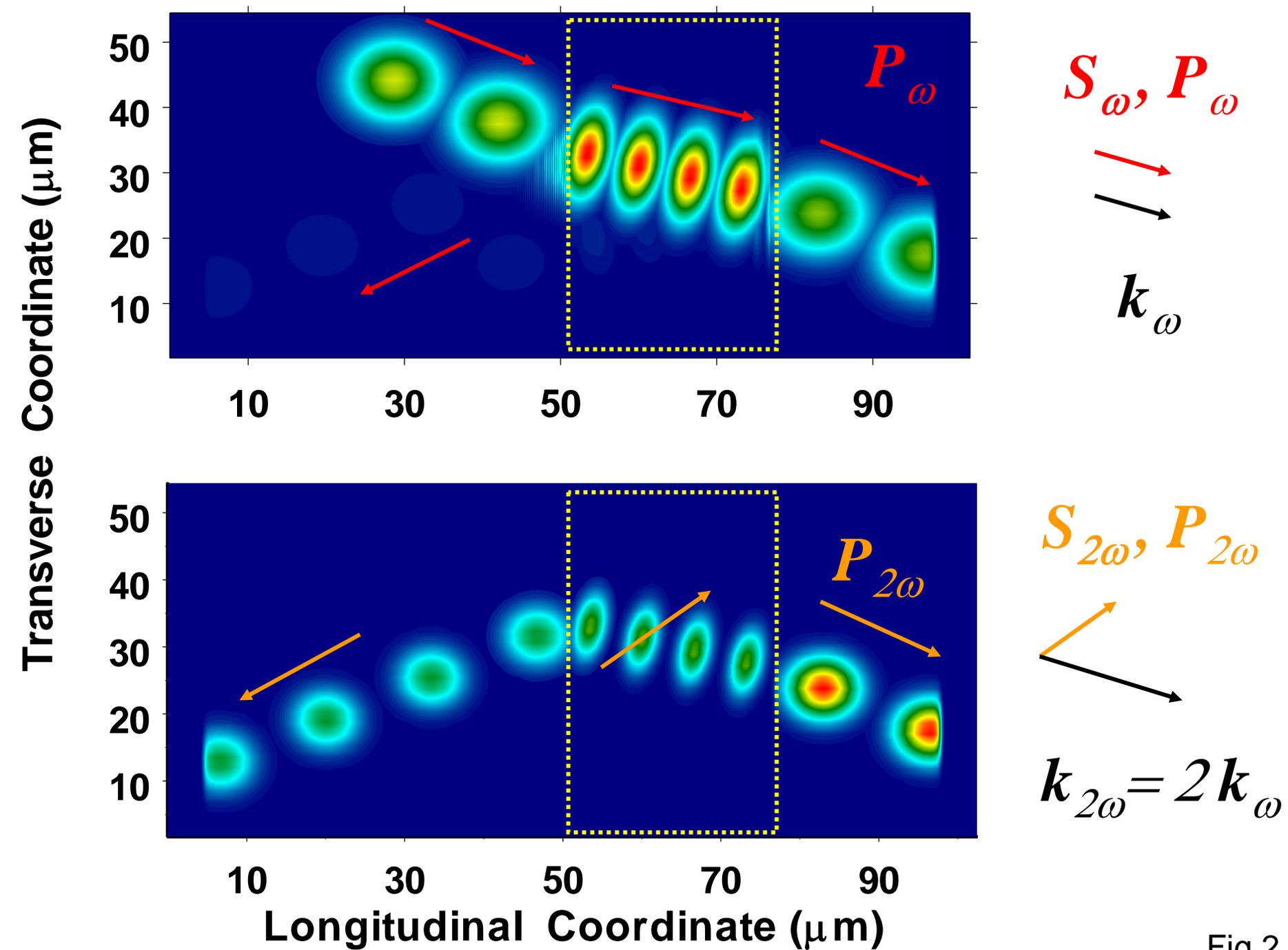

Fig.2

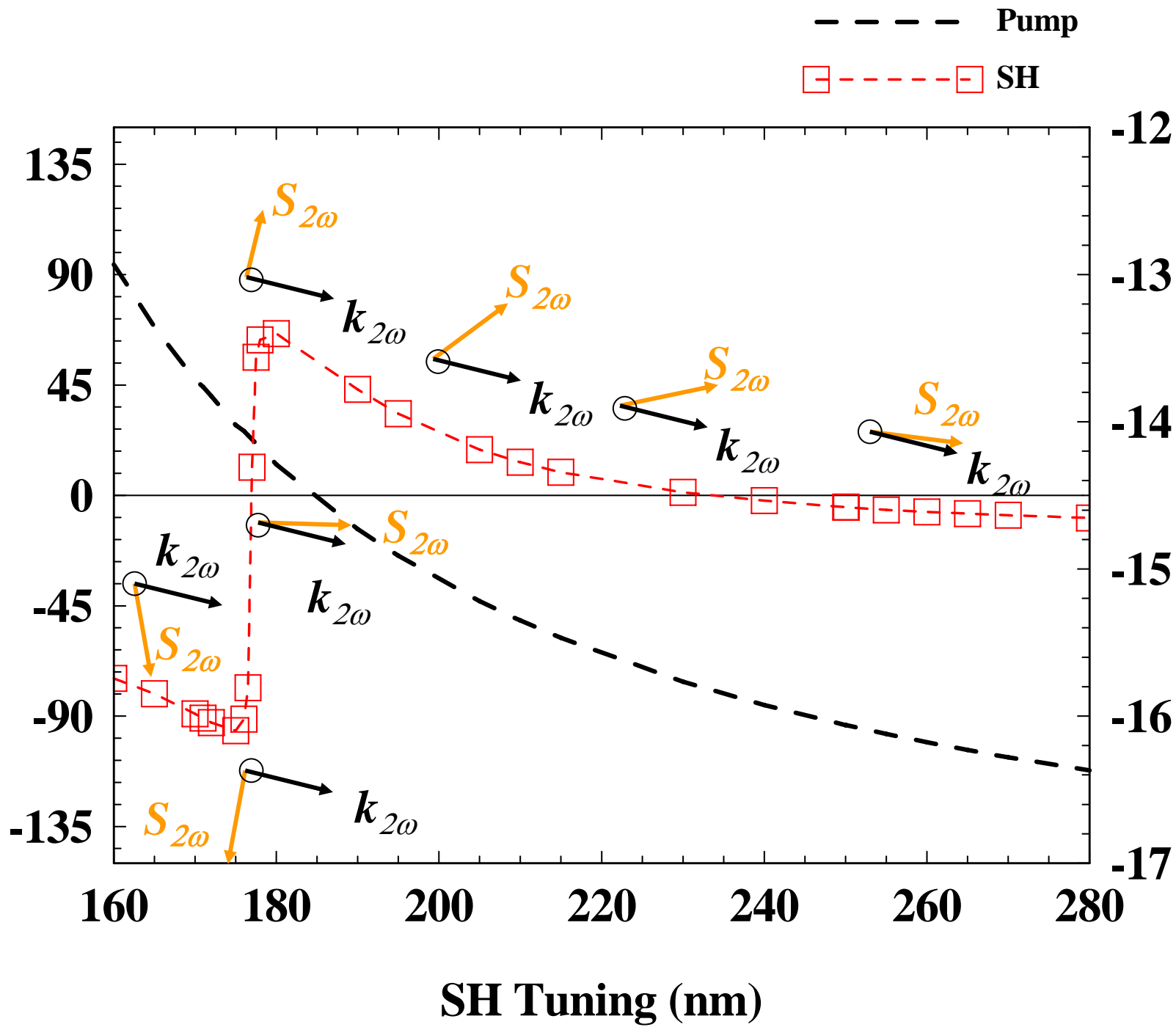

Fig.3

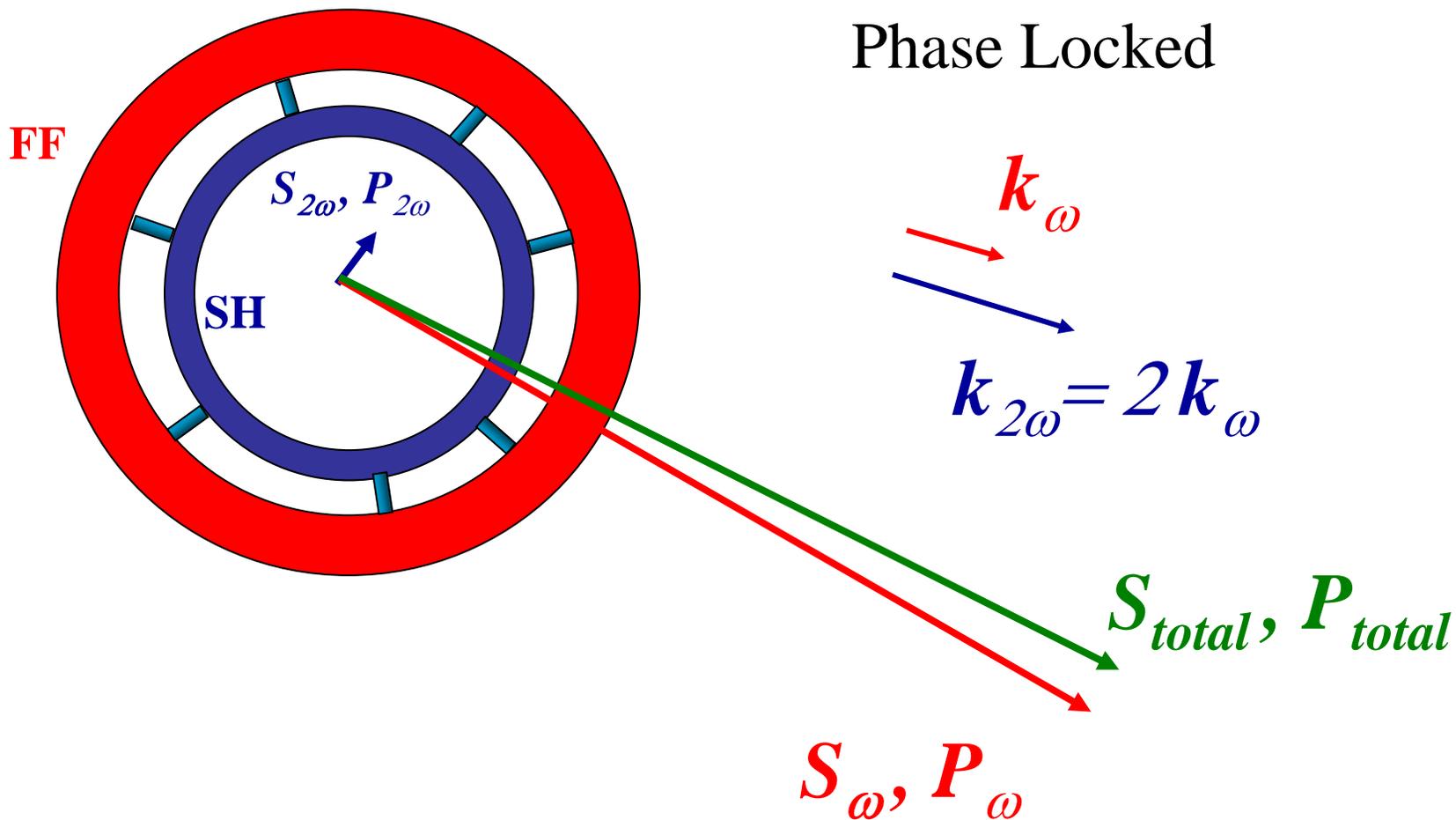

Fig.4